\documentclass[pra,showpacs,aps,twocolumn,epsfig]{revtex4}

\usepackage{psfrag,graphicx}
\usepackage{amsmath,amssymb}
\newcommand{\ket}[1]{\left| #1 \right>}
\newcommand{\bra}[1]{\left< #1 \right|}

\newcommand{\sbk}[1]{\left( #1 \right)}

\begin{document}
\title{Security of six-state quantum key distribution protocol with threshold detectors}
\author{Go Kato$^{1}$}
\author{Kiyoshi Tamaki$^{2,3}$}

\affiliation{$^{1}$NTT Communication Science Laboratories, NTT Corporation\\
3-1,Morinosato Wakamiya Atsugi-Shi, Kanagawa, 243-0198, Japan\\
$^{2}$NTT Basic Research Laboratories, NTT Corporation,\\
3-1,Morinosato Wakamiya Atsugi-Shi, Kanagawa, 243-0198, Japan\\ 
$^{3}$CREST, JST Agency, 4-1-8 Honcho, Kawaguchi, Saitama, 332-0012, Japan}
\date{\today}

\begin{abstract}
We prove the unconditional security of the six-state protocol with threshold detectors and one-way classical communication. Unlike the four-state protocol (BB84), it has been proven that the squash operator for the six-state does not exist, i.e., the statistics of the measurements cannot be obtained via measurement on qubits. We propose a technique to determine which photon number states are important, and we consider a fictitious measurement on a qubit, which is defined through the squash operator of BB84, for the better estimation of Eve's information. As a result, we prove that the bit error rate threshold for the six-state protocol ($12.611\%$) remains almost the same as the one of the qubit-based six-state protocol ($12.619\%$). This clearly demonstrates the robustness of the six-state protocol against the use of the practical devices.



\end{abstract}
\maketitle


Quantum key distribution (QKD) allows legitimated users to securely communicate, and the security of QKD, especially qubit-based QKD, has been well studied so far \cite{DLM06}. Since we have to assume any possible attack when we consider the security, the assumption of qubit-detection must be confirmed or at least its fraction must be estimated with the use of photon number resolving detectors, detector decoy idea \cite{MCL08}, or estimation method via monitoring the double click event \cite{Koashi et al}, all of which require some modifications to QKD protocols.

Another approach for the security proof of QKD with threshold detectors is to consider the so-called squash operator \cite{GLLP} which squashes an optical mode down to a qubit state. This approach only requires to assign the double-click event (detectors ``0'' and ``1'' simultaneously click) to a random bit value, which is reasonable \cite{doubleclick}. The existence of the squash operator for BB84-type measurement has been proven \cite{TT08, Beaudry08}, i.e., the statistics of the outcomes of the BB84 measurement can be interpreted as if it stemmed from the BB84 measurement on qubits whatever optical signal Bob actually receives.

One might think that the squash operator should exist for any measurement with two outcomes, including the measurement of the six-state protocol \cite{Six}, where we perform measurements along a basis, $Y$ basis, in addition to $X$ and $Z$ bases in BB84. In the case of the qubit-based six-state protocol, the measurement along the extra basis lets us learn more about Eve's information gain, resulting in a higher bit error rate threshold than that of BB84, which is a main advantage of the qubit-based six-state protocol over BB84. Unfortunately, it turns out that the squash operator for the six-state protocol is proven not to exist \cite{Beaudry08}, and it is unknown whether the advantage still holds with the use of threshold detectors.

Intuitively, sending more than one-photon is not useful for the eavesdropping since it may only increase the bit error rate, and it is hard to imagine that the advantage of the qubit-based six-state protocol suddenly vanishes once we lose information about which signal is a single-photon. In other words, to consider the security of the six-state protocol with threshold detectors is to consider the robustness of a qubit-based QKD protocol even if there is no squash operator. This is indeed one of the essential features that any practical qubit-based QKD must possess, and this issue must be seriously taken into account for the design of a qubit-based QKD protocol.

In this letter, we prove the robustness of the six-state protocol by showing the bit error rate threshold remains almost the same ($12.611\%$) compared to the one of the qubit-based six-state protocol ($12.619\%$). This result shows that sending multiple photons hardly helps Eve, which confirms the intuition mentioned above. The rate is clearly larger than the rate of BB84 with threshold detectors ($11.002\%$) \cite{TT08, Beaudry08}, and this demonstrates the advantage of using two additional states in the practical situation. We remark that our work assumes the use of a single-photon as the information carrier, but we can trivially accommodate the use of an attenuated laser source by GLLP idea \cite{GLLP}. 

This letter is organized as follows. We start with a brief description of how the protocol works, and then we move on to relatively long outlining the proof, and we devote the rest of the paper to a more detailed explanation. Finally, we summarize this letter.

Since polarization state of a single-photon and the $\frac{1}{2}$-spin state are mathematically equivalent, we use $\frac{1}{2}$-spin notation for the explanation in this letter. In the six-state protocol, Alice first generates a random bit value $b=-1, 1$ and choose one basis $\alpha$ randomly out of three bases $X$, $Y$, and $Z$. Then, she sends over a quantum channel a qubit with state being $\ket{\alpha_{b}}$ that is the eigen state of $\alpha$ basis of $\frac{1}{2}$-spin whose eigen value is $b/2$. Bob randomly chooses one basis randomly out of the three bases, and he measures the spin along the chosen direction. Alice and Bob compare over a public channel the bases they used, and keep the bit value if the bases match, othrewise discard it. Alice and Bob repeat this step many times, and they apply bit error correction \cite{nielsen} and privacy amplification \cite{nielsen} to the resulting bit string (sifted key), and they share the key.

Next, we outline our proof. Our proof employs the security proof based on complementarity scenario proposed by Koashi \cite{Koashi07}. In this proof, we consider two protocols, one is the actual protocol which Alice and Bob actually conduct, and the other one is a virtual protocol. Let us assume that Alice has a qubit state, which may be fictitious, and let $Z$ basis be Alice's key generating qubit basis. The goal of the actual protocol is that Bob agrees on Alice's bit values along $Z$ basis. On the other hand, the goal of the virtual protocol is to create an eigen state of an observable $X$, which is conjugate to $Z$, with the help of Alice and Bob's arbitrary quantum operations that commute with Alice's key generating measurement. It is proven that if Alice and Bob are free to choose which protocol to execute after the actual classical and quantum communication and if they can accomplish its goal whichever choice they have made, then unconditionally secure key can be distilled.

In order to define Alice's qubit in the six-state protocol, suppose that Alice first prepares a qubit pair in the state $\frac{1}{\sqrt{2}}\left(\ket{Z_{-1}}\ket{Z_{1}}-\ket{Z_{1}}\ket{Z_{-1}}\right)$ \cite{comment0} (we choose this singlet state to fully make use of its symmetry later), measures one of the qubit by $X$, $Y$, or $Z$-basis, and sends the other qubit to Bob. Since this process outputs the exactly the same state as the one of the actual protocol, we are allowed to work on this scenario without losing any generality. In the case that we consider the security of the key generated along $Z$ basis, and once Alice and Bob can generate $\ket{X_1}$ state in Alice's side in the virtual protocol then we are done since the agreement on the bit value in the actual protocol can be trivially made via classical error correction over a public channel (the syndrome is either encrypted \cite{Koashi05} or not \cite{TK10}). 

For the generation of $\ket{X_1}$ state, an important quantity is the so-called phase error rate, which is the ratio that Bob's estimation of Alice's bit string in $X$-basis results in erroneous, and if the estimation of the phase error rate is exponentially reliable then Alice can generate $\ket{X_1}$ by random hashing along $X$-basis \cite{BDSW, Koashi05, TK10}. More precisely, the key generation rate $G$, assuming a perfect bit error correcting code, can be expressed as $G=n_{\rm sif}\left[1-H(X)-H(Z|X)\right]$. Here, $n_{\rm sif}$ is the empirical probability of having the sifted key, $H(X)$ is Shannon entropy of the bit error, and $H(Z|X)$ is Shannon entropy of the phase error conditional on the bit error pattern. In other words, $n_{\rm sif}H(X)$ is the number of the hashing along $Z$-basis needed for the agreement of the bit values in the actual protocol and $n_{\rm sif}H(Z|X)$ is the one along $X$-basis needed for the generation of the $X$-basis eigen state in the virtual protocol \cite{Koashi07}. Hence, the key for the improvement in the key generation rate is how to minimize the conditional entropy $H(Z|X)$. 

For the estimation, we assume without loss of generality that states received by Bob are classical mixtures of photon number eigen states, and let $P_N$ be the probability of receiving a state having $N$ photons \cite{comment1}. Since we have no direct access to $P_N$, we have to assume the worst case scenario where Eve maximizes the induced phase error rate by classically mixing up each photon number state and sending them to Bob. As we will see later, it can be proven that states with photon number being greater than $3$ induces too much bit errors and we can neglect those states for the analysis. Hence, we can concentrate only on $N=1,2,3$ cases, and especially we want to derive the corresponding mutual information between the bit and phase errors.

To compute the mutual information, we introduce Bob's qubit by employing the BB84 squash operator, and we have to estimate what statistics we would have obtained if we had performed the measurement along $\tilde Y$ basis onto the resulting qubit (here, ``tilde'' means that this is about a qubit space and fictitous). In general, the actual Bob's measurement along $Y$ basis does not coincide with the measurement along $\tilde Y$ basis, however they do only when $N=1,2$ thanks to the existence of the squash operator for the six-state protocol \cite{Beaudry08}. This gives the same mutual information for $N=1,2$ as the one of the qubit-based six-state protocol. We note that to employ BB84 squash, we have to randomly pick up two bases (for the explanation, we assume that we have chosen $X$ and $Z$ bases) out of the three bases in the actual protocol. This random choice does not change the actual protocol at all. The reason is that we can always split the basis choice into two steps: the first one is the choice of two bases out of the three and then one basis is chosen from the two. Moreover, we assume in the actual protocol that Alice and Bob perform joint random bit-flip operation to make the analysis simpler.

To analyze $N=3$ case, we use the symmetry of the density operator. As a result, we can estimate the mutual information. Finally, by mixing up the photon number state $N=1,2,3$ based on the worst case scenario, we show that the bit error rate threshold for the six-state protocol with threshold detectors is $12.611\%$. This is the end of outlining the proof, and we explain why $N\ge3$ can be neglected and the derivation of the bit error rate threshold in what follows, in which we take the asymptotic limit such that the number of the pulses is infinite and we neglect statistical fluctuations.

\begin{figure}[tbp]
\begin{center}
 \includegraphics[scale=0.39]{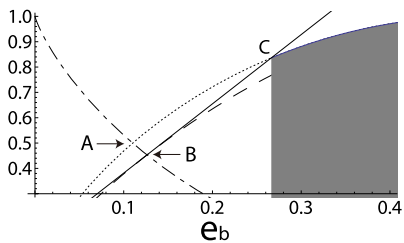}
\end{center}
 \caption{Plot of $H(Z|X)^{(1,2)}$ (dashed line), $h(e_{\rm b})$ (dotted line), $1-h(e_{\rm b})$ (dot-dashed line), and a tangent (solid line) whose tangent point ({\bf B}=(0.12619.., 0.54690..)) is the intersection of $1-h(e_{\rm b})$ and $H(Z|X)^{(1,2)}$. We can neglect any point in the gray-filled regime for the security.
\label{NP}}
\end{figure}

Our goal is to minimize $H(Z|X)$, and observe that this quantity can be rewritten as the convex combination of the conditional Shannon entropy $H(Z|X)=\sum_{N=1}^{\infty}P_{N}H(Z|X)^{\sbk N}$, where $H(Z|X)^{\sbk N}$ is the conditional Shannon entropy that is derived from $N$-photon detection event by Bob. Imagine that we make a two-dimensional ($2D$) plot of $H(Z|X)^{\sbk N}$ as a function of the bit error rate $e_{\rm b}$. The convex combination suggests that we have to consider a convex hull, each of whose extreme points corresponds to $\left(e_{\rm b}, H(Z|X)^{\sbk N}\right)$ in the $2D$ plane. Thanks to the existence of the squash operator for the six-state protocol \cite{Beaudry08}, the plot of $H(Z|X)^{(1,2)}\equiv H(Z|X)^{\sbk N}$ for $N=1,2$ is the same as the one of the qubit-based six-state protocol \cite{Six}, which is expressed as 
\begin{eqnarray}
H(Z|X)^{(1,2)}\equiv e_{\rm b}+(1-e_{\rm b})h\left(\frac{e_{\rm b}}{2(1-e_{\rm b})}\right)\,.
\end{eqnarray}
Here, $h(x)\equiv -x\log_{2}x-(1-x)\log_{2}(1-x)$, and $H(Z|X)^{(1,2)}$ is depicted in Fig.~\ref{NP} as the dashed line, in which $h(e_{\rm b})$ (dotted line), $1-h(e_{\rm b})$ (dot-dashed line), and a tangent (solid line) are also plotted. The bit error rate of the intersection ({\bf A}) of the dotted line and the dot-dashed line represents the bit error rate threshold of BB84, and the one ({\bf B}) of the dot-dashed line and the dashed line represents the bit error rate threshold of the six-state protocol up to $N=2$. {\bf C} is the intersection of the dotted line and the tangent whose tangent point is {\bf B}. Note that $H(Z|X)^{\sbk N}$ for any $N\ge3$ can never be larger than $h(e_{\rm b})$ (dotted line) as we use the squash operator for BB84. Also note that the dotted and dashed lines are concave, and an achievable point can be generated by the convex combination of a point along the dashed line and a point below the dotted line such that the average bit error rate coincides with the observed error rate. Suppose that we take convex combination of a point along the dashed line whose bit error rate is lower than the bit error rate of {\bf B} ($12.619..\%$) and a point in the gray-filled region. Since this convex combination only decreases the mutual information, it follows that we neglect any photon number state whose minimum bit error rate is larger than the bit error rate of {\bf C} ($25.677...\%$). According to analysis in \cite{appendix}, it turns out that the minimum bit error rate is strictly larger than $25.677...\%$ for $N\ge4$ (note that the minimum bit error rate is not zero for $N\ge2$ since only the singlet state ($N=1$) has the symmetry that has the zero bit error rate). Thus, we are left with working only on $N=3$ case.

\begin{figure}[tbp]
\begin{center}
 \includegraphics[scale=0.39]{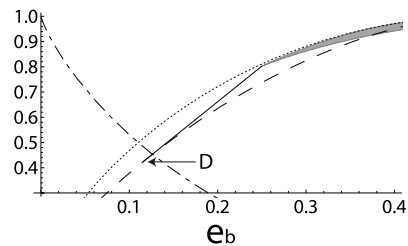}
\end{center}
 \caption{Fig.~\ref{NP} without its tangent and with $(e_b, H(Z|X)^{\sbk3})$ taking values in the shadow regime. $\overline{H(Z|X)}$ for $e_b\le0.115...$ is $H(Z|X)^{(1,2)}$ (dashed line) and the solid line represents $\overline{H(Z|X)}$ for $0.115...<e_b\le1/4$. The solid line is a tangent of the dashed line at ${\bf D}=(0.115..., 0.42407...)$, which means the slight degradation of the bit error rate threshold.
\label{NP2}}
\end{figure}

For the derivation of $H(Z|X)^{\sbk3}$, we first consider symmetrization of the state $\rho_{\rm sym}^{\sbk3}$ that Alice and Bob share. Recall that our protocol is invariant under the interchange of the basis and bit-flip in each basis. This symmetrization process is represented by a group $G$ that is generated by $\{R_{\alpha}\}$ where $R_{\alpha}$ is $\pi/2$ rotation along $\alpha$-basis ($\alpha=X, Y, Z$) of a qubit state. Also note that any rotation of the state on ${\cal H}^{\perp}$, which is an orthogonal complement to ${\cal H}$ being spanned by $\{\ket{\alpha_b}^{\otimes4}\}$, does not change the measurement outcomes since the state on ${\cal H}^{\perp}$ always induces double-click (one can also check this with POVM to be mentioned). Thus, we are allowed to work on the symmetrized density matrix $\rho_{\rm sym}^{\sbk3}\equiv\int dU\sum_{g\in G}[g\cdot \openone_{{\cal H}^{\perp}}\oplus U_{{\cal H}}]\rho_{3}[g\cdot \openone_{{\cal H}^{\perp}}\oplus U_{{\cal H}}]]^{\dagger}/|G|$. A bit tedious calculation with Shur's lemma gives us $\rho_{\rm sym}^{\sbk3}= r_0P_0+
 r_1P_1+ r_2P_2+r_3\openone_{{\cal H}^{\perp}}$ \cite{appendix}, where $r_m\geq0$ ($m=0,1,2,3$), and $P_{0,1,2}$ is a projector onto the subspace spanned by
$\{\ket{-1/2,-3/2}-\ket{1/2,3/2}$,
$\sqrt3\ket{-1/2,-1/2}+\ket{1/2,-3/2}$,
$\sqrt3\ket{1/2,1/2}+\ket{-1/2,3/2}\}$, 
$\{\ket{-1/2,1/2}-\ket{1/2,-1/2}$,
$\ket{-1/2,-1/2}-\sqrt3\ket{1/2,-3/2}$,
$\ket{1/2,1/2}-\sqrt3\ket{-1/2,3/2}\}$ and
$\{\ket{-1/2,1/2}+\ket{1/2,-1/2}$,
$\ket{-1/2,-3/2}+\ket{1/2,3/2}\}$. Here, the first (second) index in each ket represents $Z$ component of Alice's (Bob's) $\frac{1}{2}$-spin (3 $\frac{1}{2}$-spins with total angular momentum being 3/2) with eigen values being $1/2$ and $-1/2$ ($3/2$, $1/2$, $-1/2$, and $-3/2$).

To calculate the mutual information, we consider what error rate ($e_{\tilde y}$) we would have obtained if we had performed the measurement along $\tilde Y$ basis onto Alice and Bob's qubit, in which Bob's qubit is defined through the BB84 squash operator. Bob's POVM $\{M_{\alpha_b}^{\otimes4}\}$ corresponding to detection of the bit value $b=-1, 1$ along $\alpha$ basis is represented by $M_{\alpha_b}\equiv\frac12\sbk{P\sbk{\ket{\alpha_b}}^{\otimes N}-P\sbk{\ket{\alpha_{- b}}}^{\otimes N}+ \openone
}$, where $P\sbk{\ket{\alpha_b}}\equiv\ket{\alpha_b}\bra{\alpha_b}$ and $\openone/2$ represents the random assignment of the double-click event, and POVM for detecting $\alpha$-basis error $\Gamma_\alpha$ is \cite{comment0} $\Gamma_\alpha\equiv
P\sbk{\ket{\alpha_1}}\otimes M_{\alpha_1}+
P\sbk{\ket{\alpha_{-1}}}\otimes M_{\alpha_{-1}}$.
POVM for detecting $\tilde Y$ basis error on {\it qubit pair} is given by $\tilde\Gamma_y\equiv
P\sbk{\ket{Y_1}}\otimes{\cal F}_{\rm BB84}\left(P\sbk{\ket{Y_1}}\right)+P\sbk{\ket{Y_{-1}}}\otimes{\cal F}_{\rm BB84}\left(P\sbk{\ket{Y_{-1}}}\right)$, where ${\cal F}_{\rm BB84}(\cdot)$ is a map from the qubit space to 3-photon space, which is represented by Kraus operator for the BB84 squash \cite{TT08, Beaudry08}. Using all of them, the bit error rate $e_b$ and $e_{\tilde y}$ are respectively represented by $e_b\equiv{\rm Tr}\Gamma_{\alpha} \rho_{sym}^{\sbk{3}}=\frac14 r_0-\frac34r_1+\frac12r_2+\frac12$ and $e_{\tilde y}\equiv{\rm Tr}\tilde\Gamma_y \rho_{sym}^{\sbk{3}}=\frac12
   r_0-\frac12r_1+\frac12$, and what we have to do is to derive $e_{\tilde y}$ as a function of $e_{b}$ and to maximize $ H(Z|X)^{\sbk3}$. In the equation of $e_{b}$ and $e_{\tilde y}$, we erase the parameter $r_3$ by using the condition ${\rm Tr}\rho_{sym}^{\sbk3}=1$, which follows that the positivity condition of $\rho_{sym}^{\sbk 3}$ reads $r_0,r_1,r_2\geq0$ and $3r_0+3r_1+2r_2\leq1$. By introducing a parameter set $\{t, s, u\}$ with $0\le t, u\le1$ and $-1\le s\le1$, we can express $r_0=ut(1+s)/6$, $r_1=ut(1-s)/6$, and $r_2=u(1-t)/2$, and we use this parameterization to derive the regime $\{e_b, e_{\tilde y}\}$ that $\rho_{sym}^{\sbk 3}$ can take. The regime is represented by the triangle with vertices being $\{1/4, 1/3\}$, $\{7/12, 2/3\}$, and $\{3/4, 1/2\}$ in $\{e_b, e_{\tilde y}\}$-plane, which means that $e_{\tilde y}$ is always bounded by linear functions of $e_b$. This triangle can be translated into the shadow regime in Fig.~\ref{NP2} via $H(Z|X)^{(3)}=e_{b}h[(2e_{b}-e_{\tilde y})/(2e_b)]+(1-e_{b})h[e_{\tilde y}/(2-2e_b)]$ that coincides with $H(Z|X)^{(1,2)}$ when $e_b=e_{\tilde y}$, and we note that the tangent in Fig.~\ref{NP} crosses the shadow regime in Fig.~\ref{NP2} so that the bit error rate threshold should degrade. By considering the convex hull of $H(Z|X)^{\sbk N}$ for $N=1,2,3$, the upper bound of $H(Z|X)$, which we express as $\overline{H(Z|X)}$, is given by 
\begin{eqnarray}
&& \overline{H(Z|X)}
\nonumber\\
&=&\left\{
\begin{array}{cl}
 H(Z|X)^{(1,2)}
&\makebox{ in case  $0.115...>e_b$} \\
\sbk{2.82...}e_b+0.0976...&\makebox{ in case $\frac14\geq e_b\geq0.115...$} \\
\end{array}
\right.
\nonumber
\end{eqnarray}
This is also shown in Fig.~\ref{NP2}. From this expression, we can derive the bit error rate threshold of $12.6112...$\% by solving $\overline{H(Z|X)}=1-h\sbk{e_b}$ with respect to $e_b$.

{\it Remarks:}
For the first sight, our analysis assumes that Alice and Bob's pair states are identically and independently distributed. A way to treat unconditional security is to use the argument based on quantum de Finetti theorem \cite{Finetti} or Azuma's inequality \cite{A67, Azuma}. In the latter argument, we consider an arbitrary whole Alice and Bob's state, not just a pair state, and we consider to perform the Bell basis measurement from the first qubit pair in order. $\rho_{\rm sym}$ is now interpreted as the state of a particular qubit pair conditional on arbitrary Bell basis measurement outcomes. It follows that $e_{\tilde y}$ and $e_b$ are probability also being conditional on the outcomes, which is a property required in applying Azuma's inequality, and most importantly the relation between them are linear as we have already mentioned (for $4\le N$ case, it is given by $0\le e_{\tilde y}\le1$ and $0.25677...\le e_b$). Thus, we can convert our analysis into the analysis of the unconditional security proof by using exactly the same argument as \cite{Azuma}. 

To summarize, we prove the unconditional security of the six-state protocol with threshold detectors. For the proof, we propose a technique to determine which photon number states are important, and we employ the squash operator for BB84 and the estimation of the mutual information that can be obtained via $Y$ basis fictitious measurement on the resulting qubit state. In this letter we consider one-way quantum communication protocol, and our analysis may apply to two-way quantum communication protocol such as BBM92 type QKD \cite{BBM92}, which we leave for the future study. Security proof of other protocols with threshold detectors are also another future works. 

We thank Hoi-Kwong Lo, Marcos Curty, and Koji Azuma for valuable comments and discussions. This work was in part supported by National Institute of Information and Communications Technology (NICT) in Japan.


\begin{thebibliography}{}
\bibitem{comment0} Note that Alice and Bob's bit values are anti-correlated with this state in the absence of Eve.
\bibitem{comment1} This is because introducing projection onto Fock space preceding Bob's measurement does not change any measurement outcome.
\end{thebibliography}
\end{document}